\documentclass[12pt,a4paper]{article}

\usepackage{latexsym}
\usepackage{graphicx,color}
\usepackage{amsmath}
\newcommand\sect[1]{\section{#1}\setcounter{equation}0}
\newcommand\ds{\displaystyle}
\newcommand\no{\nonumber\\{}}
\newcommand\eqnb{\begin{eqnarray}}
\newcommand\eqne{\end{eqnarray}}

\usepackage{latexsym}
\begin{document}
\center{{\huge The BRST treatment of stretched membranes}\vspace{15mm}\\
Jonas Bj\" ornsson\footnote{jonas.bjornsson@kau.se} and Stephen 
Hwang\footnote{stephen.hwang@kau.se}\\Department of Physics\\Karlstad University
\\SE-651 88 Karlstad, Sweden}\vspace{15mm}\\

\abstract{
The BRST-invariant formulation of the bosonic stretched membrane is considered. In this
formulation the stretched membrane is given as a perturbation around zero-tension 
membranes, where the BRST-charge decomposes as a sum of a string-like BRST-charge and
a perturbation. It is proven, by means of cohomology techniques, that there exists
to any order in perturbation theory
a canonical transformation that reduces the full BRST-charge to the string-like one.
It is also shown that one may extend the results to the quantum level yielding a 
nilpotent charge in 27 dimensions.}

\newpage

\sect{Introduction}
Membranes are interesting from many points of view, it may have a connnection to M-theory 
\cite{Townsend:1995kk,Witten:1995ex} and it is a generalization of the string action. In the 
lightcone gauge it can be reduced to a matrix model \cite{goldstone,hoppe,deWit:1988ig} which is 
conjectured to be M-theory \cite{Banks:1996vh}. It is probably also relevant as a D$2$-brane, being
part of the strongly coupling region of string theory. The relevance of D-branes 
for string theory at strong coupling was first realized in \cite{Polchinski:1995mt}. 
It is also interesting by itself and 
as a testing ground to see if methods in string theory generalize to higher extended objects. 
Solutions of the equations of motions are rare because of the highly non-linear equations of 
motion. 

In \cite{Bjornsson:2004yp} we proposed to study so-called stretched membrane configurations. 
These are configurations which arise, in a partial fixing of the gauge, for weak tensions of the 
membrane. 
Stretched membranes may be treated perturbatively around a zero tension limit, which corresponds 
to a string-like theory.
In \cite{Bjornsson:2005rv} we proved that, by fixing the gauge completely to the lightcone gauge, 
there is a canonical equivalence between the two theories i.e.\ the membrane is, to any order in perturbation 
theory, equivalent to a string-like theory. Properties of stretched membranes may, therefore, be inferred 
from those of the string-like theory.
The equivalence holds for bosonic as well as supersymmetric membranes. It was also shown that
the canonical equivalence extends to a unitary one at the quantum level, yielding, among other results,
the critical dimensions 27 and 11 for the bosonic and supersymmetric cases, respectively.

In this article we continue our analysis of the bosonic stretched membrane. The aim is to see whether 
the 
equivalence may be proven without the use of the light-cone gauge. Classically, one may argue that 
this must
be the case, at least locally in phase-space. But at the quantum level this need not be true. 
Proving unitary equivalence for a fully gauged
fixed theory does not, in general, imply that the same is true without gauge fixing, since the gauge
symmetry may break down due to anomalies. It is rather the converse that
is true. By proving the unitary equivalence between the BRST-charge of the stretched membrane
and the unperturbed string-like BRST-charge, we can conclude that, since the latter theory is 
non-anomalous, this 
is also true for the former one. From this it follows 
that we can impose any particular gauge and still maintain equivalence.

The problem to solve is, therefore, the following. Given a BRST-charge of the form
\eqnb
Q=Q^0+Q'\label{Qmem},
\eqne
where $Q^0$ is the unperturbed string-like BRST-charge and $Q'$ the perturbation, is it possible to 
find a canonical transformation which 
takes $Q$ into $Q^0$? Unfortunately, the techniques used in \cite{Bjornsson:2005rv}
do not readily generalize to the present case due to the complexity of the problem. 
Instead, we will use another approach. As we will see, it is possible to restate the problem as
one of the cohomology of $Q^0$. If this cohomology is trivial for ghost number one then 
there will always exist, to any order in perturbation theory, a canonical transformation of the kind
we are looking for. In fact, as we will show, the restatement of the perturbation problem as one
in cohomology is not something particular for stretched membranes, but is quite general.

Since $Q^0$ is essentially the BRST-charge of the bosonic string, the cohomology problem seems already to  
have been solved. This is not entirely correct. 
First of all, the basic fields including the ghosts, are fields defined
on the world-volume rather than the world-sheet. Secondly, the known proofs of the cohomology of string theory 
do not directly apply to our case. 
The cohomology w.r.t.\ the quantum string state-space is well known \cite{Kato:1982im, 
Freeman:1986fx, Frenkel:1986dg, Thorn:1986xy, Polchinski:1998rq}. Using the one-to-one 
correspondence between operators and states, one may also deduce the cohomology of the operators. 
In our case, we need to analyze the classical cohomology of the phase-space functions, which turns out 
to be a little bit different.

Having established the canonical equivalence one can turn to the quantum case. We will show that the 
quantization procedure proposed in
\cite{Bjornsson:2005rv} can, in a straightforward way, be applied to the present case. By this 
procedure one defines a specific ordering whereby the canonical transformations turn 
into unitary ones so that the equivalence of the perturbed and unperturbed theories 
is maintained at the quantum level. This then will show that quantum consistency, through the
nilpotency of the membrane BRST-charge, requires 27 dimensions. 

The paper is organized as follows.
In section two we consider the BRST treatment of gauge theories formulated as perturbation theories. Here  
we also show the connnection between the existence of canonical transformations and the BRST cohomology. 
In the third section we discuss the 
cohomology of the BRST-charge locally in phase-space. The cohomology problem relevant to the membrane is 
treated in section four. 
This will also show the canonical equivalence between the string-like theory and the stretched membrane. 
In the last section we discuss the quantization of our model. 

%%%%%%%%%%%%%%%%%%%%%%%%%%%%%%%%%%%%%%%%%%%%%%%%%%%%%%%%%

\sect{BRST treatment of perturbatively formulated gauge theories}

In this section we will, in more general terms, formulate the problem of finding canonical 
transformations which canonically map constraint theories formulated as an unperturbed and a 
perturbed part. We start with a general theory of this form. In our particular case, the 
unperturbed theory is a 
string-like theory, which is the standard bosonic string theory with extra 
world volume parameter dependence. The perturbed theory is the stretched membrane theory formulated 
in \cite{Bjornsson:2004yp}.

Consider the general situation where we have a theory with first-class constraints, $\phi_a\approx 0$,
formulated as a perturbation theory
\eqnb
\phi_a[p_i,q^i]&=&\psi_a[p_i,q^i]+\sum_{n=1}^{N}g^n\lambda_a^{(n)}[p_i,q^i]
\label{perturbation theory}
\eqne
where $g\ll 1$ is the perturbation parameter. Since we have a closed Poisson bracket algebra 
\eqnb
\left\{\phi_a,\phi_b\right\}={U_{ab}}^c\phi_c
\eqne
to any order in $g$, it follows that the unperturbed part, $\psi_a$, also satisfies a closed 
algebra
\eqnb
\left\{\psi_a,\psi_b\right\}={U'_{ab}}^c\psi_c,
\eqne
where, in general, ${U_{ab}}^c$ and ${U'_{ab}}^c$ can depend on the phase-space variables and 
\eqnb
{U'_{ab}}^c\equiv{U_{ab}}^c\mid_{g=0}.
\eqne
The BRST-charge is generally of the form
\eqnb
Q=\sum_{n=0}\stackrel{(n)}{Q},
\eqne
where
\eqnb
\stackrel{(0)}{Q}&=&\phi_a c^a\no
\stackrel{(n)}{Q}&=&A_{a_1,\dots,a_{n+1}}^{b_1,\dots,b_n}c^{a_1}\cdot\ldots\cdot
c^{a_{n+1}}b_{b_1}\cdot\ldots\cdot b_{b_n},
\eqne
and the functions, $A_{a_1,\dots,a_{n+1}}^{b_1,\dots,b_n}$, are determined by the Poisson bracket algebra 
of the constraints and the nilpotency condition on the BRST-charge. 

In the assumed perturbation theory we can also expand in terms of $g$
\eqnb
Q&=&Q^0+\sum_{i=1}^{N'}g^iQ^i.
\eqne
The nilpotency condition of the full BRST-charge now yields relations to each order
in $g$. The zeroth order relation gives that $Q^0$, the BRST-charge for the unperturbed theory, is nilpotent.
To first order in $g$ we have
\eqnb
\{Q^0,Q^1\}&=&0\label{Qfirst}.
\eqne
Thus, we know that $Q^1$ is in the cohomology of $Q^0$. If $Q^1$ is a trivial element in this
cohomology then there exists a function $G_1$ satisfying
\eqnb
Q^1&=&-\{Q^0,G_1\}.
\label{solution of Q1}
\eqne
Let us assume that $G_1$ exists. Then we are free to interpret 
$G_1$ as a generator of an infinitesimal canonical transformation. This transformation shifts $Q$ 
to 
\eqnb
Q&\stackrel{G_1}{\longrightarrow}&Q^0+g^2\left(Q^2-\frac{1}{2}\left\{\left\{Q^0,G_1\right\},
G_1\right\}\right)+\dots.
\eqne
The nilpotency condition to second order in $g$ is
\eqnb
\left\{Q^0, Q^2-\frac{1}{2}\left\{\left\{Q^0,G_1\right\},G_1\right\}\right\}&=&0.
\eqne
This implies that ${Q'}^2\equiv Q^2-\frac{1}{2}\left\{\left\{Q^0,G_1\right\},G_1\right\}$ is in the 
cohomology
of $Q^0$ and we have a problem of the same type as in eq.\ (\ref{Qfirst}). If ${Q'}^2$ is a trivial 
element in the 
cohomology, then we may repeat the above argument and conclude that there exists an infintesimal 
canonical transformation that
transforms $Q$ to $Q^0$ to second order in $g$. One may continue this to any order in $g$. Thus, we see that
the problem of proving that there exists a canonical transformation to any order in perturbation theory may be
solved by proving that the $Q^0$ cohomology at ghost number one is trivial.

It should be remarked that the above argument goes through for the quantum case as well. Replacing all Poisson brackets
with commutators shows that the problem of unitary equivalence can be restated in terms of the cohomology of the
BRST-operator. For the stretched membrane, however, this is not helpful. The argument requires that
one can establish the nilpotency at the quantum level of the BRST-operator and this we cannot do from the outset.
Instead we will have to proceed through the classical analysis and, using this, define 
a quantum theory by promoting the canonical transformations to unitary ones. The nilpotency will then follow 
as a consequence of the unitary equivalence. 

%%%%%%%%%%%%%%%%%%%%%%%%%%%%%%%%%%%%

\sect{Local existence of a canonical transformation}

In this section we will continue to consider the general situation, but only locally in phase-space.
We will show that in this case there always exists a canonical transformation of the type discussed above.

The starting point is the again a theory as in section two with constraints $\phi_a$.
We will here first use the abelization theorem \cite{Batalin} (for a short proof of it, see 
\cite{Henneaux:1992ig}). The 
theorem states that for all constraint theories there exists, locally in phase-space, an invertible 
coordinate dependent matrix 
${K_a}^b$ such that $F_a \equiv {K_a}^b\phi_b$ are 
abelian. For explicit constructions for the free bosonic string theory, see \cite{Aratyn:1987nw,Hwang:1990mx}. 
A theorem by Henneaux \cite{Henneaux:1985kr} shows that there exists a canonical 
transformation, $G$, in the extended phase-space such that the BRST-charge of the 
unperturbed theory is canonically equivalent to an abelian one,
\eqnb
Q^0&\stackrel{G}{\longrightarrow}&\tilde Q^0=F_ac^a.
\label{tilde Q^0}
\eqne
Applying this canonical transformation to the full theory yields a BRST-charge 
\eqnb
Q\stackrel{G}{\longrightarrow}\tilde Q&=&\tilde Q^0+\sum_{i=1}^{N'}g^i\tilde Q^i\label{tilde Q}
\eqne
where $\tilde Q^0$ is given by eq.\ (\ref{tilde Q^0}).

As we discussed in the previous section, the existence of a canonical transformation, which 
maps the perturbed BRST-charge to the unperturbed one, is determined by the cohomology of the 
unperturbed BRST-charge. Let us, therefore, study the cohomology of the simple abelian model 
in more detail. 

Assume that there exist $m$ first class constraints, $F_a \approx 0$, in a theory with 
$n$ degrees of freedom. Since the theory is abelian there exists locally, by Darboux's theorem, a canonical 
transformation from $(q^i,p_i)$ to $(\chi^a,{q^*}^j,F_a,{p^*}_j)$, where $j=1,\ldots,n-m$ and 
$\{\chi^a,F_b\}=\delta^a_b$. The BRST-charge for the abelian model is
\eqnb
Q_{A}=F_ac^a.
\eqne
One can also add ghost momenta, $b_a$, that satisfy 
\eqnb
\left\{c^a,b_b\right\}&=&\delta^a_b.
\eqne
We will restrict our study of the BRST cohomology to the 
space of polynomials in the phase-space coordinates. The proof we will give 
will not depend on the assumption of locality. This will be important as we will need
the result in the next section, where the treatment is not restricted to being
local.

Let us construct $m$ charges from $Q_A$ (no summation over $a$)
\eqnb
N_a&\equiv&\left\{Q_A,\chi^ab_a\right\}\no
&=&\chi^aF_a-c^ab_a.
\eqne
A non-trivial BRST-invariant function has to have zero eigenvalues, in the 
Poisson bracket sense, w.r.t.\ any of these charges. Otherwise, if
a BRST-invariant function $\mathcal O$ satisfies
\eqnb
\{N_a,\mathcal O\}&=&n_a \mathcal O,
\eqne
it is BRST-trivial
\eqnb
\mathcal O&=&\frac{1}{n_a}\{Q_A,\{\mathcal O,\chi^ab_a\}\}.
\eqne
The fundamental fields in this theory with non-zero eigenvalues of $N_a$ are $(F_a,b_a)$ with 
eigenvalue $+1$ and $(\chi^a,c^a)$ with eigenvalue $-1$. Thus, non-trivial BRST-invariant 
polynomials can depend on $(\chi^a,F_a,c^a,b_a)$ only through the combinations
$(F_ac^a,F_a\chi^a,b_ac^a,b_a\chi^a)$ 
(no summation over $a$). 
Define these linear combinations (no summation over $a$) 
\eqnb
s_a&\equiv&\frac{1}{2}b_a\chi^a\no
t_a&\equiv&\frac{1}{2}\left(b_ac^a+F_a\chi^a\right)\no
u_a&\equiv&\frac{1}{2}\left(b_ac^a-F_a\chi^a\right)\no
v_a&\equiv&F_ac^a.\label{defs-v}
\eqne
They satisfy (no summation over $a$) 
\eqnb
&&s_a\stackrel{Q_A}{\longrightarrow} t_a\stackrel{Q_A}{\longrightarrow} 0\\
&&u_a\stackrel{Q_A}{\longrightarrow} v_a\stackrel{Q_A}{\longrightarrow} 0\\
&&s_a^2=v^2_a=0\label{s1}\\
&&t_a^2-u_a^2=2s_av_a\label{s2}\\
&&(u_a+t_a)v_a=0. \label{(t-u)v}
\eqne
Eqs. (\ref{s1}) and (\ref{s2}) imply that we can reduce any polynomial to be at most 
linear in $s_a$, $v_a$ and $u_a$. 

Let us determine the cohomology of the BRST-charge by first fixing to a generic value of
$a$ and suppress the indices of the fields $(s_a,v_a,u_a,t_a)$. Let $f(s,t,u,v,x)$ be a 
BRST-invariant function where $x$ indicates dependence on other fields. Expand first
the $s$-dependence of $f$
\eqnb
f&=&sf_1(t,u,v,x)+f_2(t,u,v,x).
\label{expansion of f nr1}
\eqne
The BRST-invariance of $f$ implies
\eqnb
\{Q_A,f_1\}&=&0\\
\{Q_A,f_2(t,u,v,x)\}+tf_1(t,u,v,x)&=&0.
\eqne
The second equation implies that $f_2$ can be split into two parts
\eqnb
f_2&=& tG_1(t,u,v,x)+f_3(t,u,v,x)
\label{f2}
\eqne
where
\eqnb
\{Q_A,G_1(t,u,v,x)\}&=&-f_1\no
\{Q_A,f_3(t,u,v,x)\}&=&0\nonumber.
\eqne
This is always possible, because otherwise $f_1=0$. 
Inserting eq.\ (\ref{f2}) into eq.\ (\ref{expansion of f nr1}) yields
\eqnb
f&=&-s\{Q_A,G_1\}+tG_1(t,u,v,x)+f_3(t,u,v,x)\no
&=&\{Q_A,sG_1\}+f_3(t,u,v,x).
\eqne
Thus, non-trivial functions in the cohomology of $Q_A$ are independent of $s$. We can 
expand $f_3$ as
\eqnb
f_3&=&uvf^1_4(t,x)+uf^2_4(t,x)+f_5(t,v,x).
\eqne
The BRST-invariance of $f_3$ implies
\eqnb
\{Q_A,f^1_4(t,x)\}&=&0\no
\{Q_A,f^2_4(t,x)\}&=&0\no
\{Q_A,f_5(t,v,x)\}+vf^2_4(t,x)&=&0
\label{f5 and f24}.
\eqne
The first equation shows us that $uvf_4^1$ is trivial
\eqnb
\left\{Q_A,vsf^{1}_4\right\}&=&-tvf^1_4,\no
&=&uvf^1_4,
\eqne
where the last equality follows from eq.\ (\ref{(t-u)v}).
Eq.\ (\ref{f5 and f24}) also implies that one can split $f_5$ into 
two parts
\eqnb
f_5&=&vG_2(t,x)+f_6(t,v,x),
\eqne
with
\eqnb
\{Q_A,G_2\}&=&f^2_4\no
\{Q_A,f_6\}&=&0.
\eqne
Extracting the $v$-dependence of the function $f_6$, 
\eqnb
f_6&=&vf_7(t,x)+f_8(t,x),
\eqne
shows us that the BRST-invariance of $f$ implies that both functions, $f_7$ and $f_8$, are 
BRST-invariant.
The first term is BRST-trivial, $vf_7=\{Q_A,uf_7\}$. Expanding the $t$-dependence of $f_8$
\eqnb
f_8&=&\sum_{j=0}^{\infty}t^jh_j(x),
\eqne
yields that each $h_j$ has to be BRST-invariant. This implies that
\eqnb
f_8= \{Q_A,G_3\}+h_0(x),
\eqne
where
\eqnb
G_3&=&\sum_{j=1}^{\infty}st^{j-1}h_j(x).
\eqne
Collecting all parts we have
\eqnb
f&=&\left\{Q_A,sG_1+uG_2+vsf^1_4+uf_7+G_3\right\}+h_0(x).
\eqne
Concluding, we have shown that all non-trivial phase-space polynimials in the cohomology 
are independent of $t_a$, $u_a$, $v_a$ and $s_a$, for a fixed value of $a$. This is true for all 
values of $a$. Thus, the only non-trivial elements in the cohomology are ghost number zero polynomials
that only depend on $({q^*}^j,{p^*}_j)$, the coordinates that span the 
physical phase-space. 

Let us now return to our problem of proving the existence of a canonical transformation. From our 
results of the cohomology and from the previous section we have proven that 
such a transformation exists to all orders in perturbation theory. This implies that 
we have proven the existence of the canonical transformations $G$ and $G'$ such that
\eqnb
\begin{array}{ccccccc}
\ds Q&\ds \stackrel{G}{\longrightarrow}&\tilde Q&\ds \stackrel{G'}{\longrightarrow}&\tilde Q^0
&\ds \stackrel{G^{-1}}{\longrightarrow}&Q^0,
\end{array}
\eqne
where $G$ transforms the unperturbed constraints to abelian ones, $G'$ transforms the perturbed BRST-charge to the 
unperturbed abelian one and, finally, $G^{-1}$ transforms us to the original unperturbed BRST-charge.
These statements are true locally. There may still, however, exist obstructions preventing 
the results to hold globally.

%%%%%%%%%%%%%%%%%%%%%%%%%%%%%%%%%%%%%%%%%%%%%%%%%%%%%%%%%

\sect{Application to the stretched membrane}

We have seen from the previous section that
we are assured that there exists, at least locally, a canonical transformation transforming the 
full BRST-charge to the unperturbed one. We will now consider what happens
in the specific case of the stretched membrane when we do not restrict ourselves to local
considerations. Although the results of the local case will not be needed as such, we do need to use the
result from the analysis of the abelian BRST-charge, which was not restricted to be local in phase-space.

For the stretched membrane the unperturbed BRST-charge is of the form eq.\ (\ref{Qmem}) where $Q^0$ is that 
of a free string with an extra world-parameter dependence. The cohomology of the 
state-space of the free quantum string theory is well known, see \cite{Kato:1982im, 
Freeman:1986fx, Frenkel:1986dg, Thorn:1986xy,Polchinski:1998rq}. The cohomology of the 
classical theory 
has, however, not to our knowledge been solved. Using techniques largely based on 
\cite{Polchinski:1998rq}, we will analyse the cohomology of our string-like model. 

If we reduce one of the three constraints for the membrane theory and introduce two ghosts 
and ghost momenta for the remaining constraints, one can construct a BRST-charge for this theory\footnote{We have 
corrected a sign error in \cite{Bjornsson:2004yp}}
\eqnb
Q&=&\int d^2\xi \mathcal Q
\eqne
\eqnb
\mathcal Q&=&\phi_1c^1+\phi_2c^2+\partial_1 c^1c^1b_1+\partial_1c^2c^2b_1+
\partial_1c^2c^1b_2+\partial_1c^1c^2b_2\no
&+&
g\left[\mathcal P\partial_2 X\partial_2c^1c^2b_1
-\partial_1 X\partial_2 X\partial_2c^2c^2b_1
+\left(\partial_2 X\right)^2\partial_1c^2c^2b_1
\right.
\no
&+&
\left.
2\mathcal P \partial_2 X\partial_2c^2c^2b_2-2\partial_2 c^1\partial_2 c^2c^2b_1b_2\right],
\label{brst charge}
\eqne
where
\eqnb
\phi_1&=&\mathcal P \partial_1 X\no
\phi_2&=&\frac{1}{2}\left\{\mathcal P^2+\left(\partial_1 X\right)^2\right.\no
&+&
\left.
g\left[
\left(\partial_1 X\right)^2\left(\partial_2 X\right)^2+\left(\mathcal P 
\partial_2 X\right)^2-\left(\partial_1 X \partial_2 X\right)^2\right]\right\}.
\eqne
We can split this BRST-charge into two parts, one free part, which is that of a string-like 
theory, and a perturbation
\eqnb
Q&=&Q^0+gQ^1.
\eqne
If we make a change of variables from $(X^\mu,\mathcal P_\mu)$, $\mu=0,\ldots D-2$, to the Fourier coefficients 
$(q^\mu_n,\tilde q^\mu_n,\alpha^\mu_{m,n},\tilde \alpha^\mu_{m,n})$,
we have the non-zero Poisson brackets 
\eqnb
\{\alpha^\mu_{m,n},\alpha^\nu_{p,q}\}&=&\{\tilde \alpha^\mu_{m,n},\tilde \alpha^\nu_{p,q}\}
=-im\eta^{\mu\nu}\delta_{m+p,0}\delta_{n+q,0}\no
\{q_m^\mu,\alpha_{0,n}^\nu\}&=&\{\tilde q_m^\mu,\tilde\alpha_{0,n}^\nu\}=\eta^{\mu\nu}\delta_{m+n,0}
\eqne
To simplify the equations, we redefine our ghosts and ghost momenta
\eqnb
c&=&c^1+c^2\no
\tilde c&=&c^1-c^2\no
b&=&\frac{1}{2}\left(b_1+b_2\right)\no
\tilde b&=&\frac{1}{2}\left(b_1-b_2\right).
\eqne
Fourier expanding these fields we find the non-zero Poisson brackets
\begin{equation}
\begin{array}{rcccl}
\{c_{m,n},b_{p,q}\}&=&\{\tilde c_{m,n},\tilde b_{p,q}\}&=&\delta_{m+p,0}\delta_{n+q,0}.
\end{array}
\end{equation}
Choose lightcone coordinates
\eqnb
A^+&=&\frac{1}{\sqrt{2}}\left(A^{D-2}+A^{0}\right)\no
A^-&=&\frac{1}{\sqrt{2}}\left(A^{D-2}-A^{0}\right)
\eqne
and introduce a grading by
\eqnb
N_{lc}
&=&
\sum_{m\neq 0,n}\frac{1}{im}\left(\alpha^+_{-m,-n}\alpha^-_{m,n}
+\tilde\alpha^+_{-m,-n}\tilde\alpha^-_{m,n}\right)
+\sum_{n\neq 0}\left(\alpha^-_{0,-n}q^+_n+\tilde\alpha^-_{0,-n}\tilde q^+_n
\right.
\no
&-&
\left.
\alpha^+_{0,-n}q^-_n-\tilde\alpha^+_{0,-n}\tilde q^-_n\right)
+\alpha^-_{0,0}q^+_0+\tilde\alpha^-_{0,0}\tilde q^+_0.
\label{N{lc}}
\eqne
$N_{lc}$ acts diagonally, within Poisson brackets, on the basic fields.  $q^-_{n\neq0}$, 
$\tilde q^-_{n\neq0}$, $\alpha^-_{m,n}$ 
and $\tilde\alpha^-_{m,n}$ have eigenvalue $+1$; $q^+_{n}$, $\tilde q^+_{n}$ for all $n$,  
$\alpha^+_{m,n}$ and $\tilde \alpha^+_{m,n}$ for $\left|m\right|+\left|n\right|\neq0$ have eigenvalue $-1$. All other 
fields have eigenvalue zero.

The string-like BRST-charge may now be split into two parts
\eqnb
Q^0&=&Q_1+Q_0,
\eqne
where the lower index indicates the eigenvalue w.r.t.\ $N_{lc}$. The nilpotency of
$Q^0$ implies
\eqnb
\{Q_1,Q_1\}=\{Q_0,Q_1\}=\{Q_0,Q_0\}=0.
\eqne
Thus, the two separate terms in $Q^0$ are nilpotent by themselves. 
The explicit form of $Q_1$ is simple
\eqnb
Q_1&=&\sum_{m,n}\left(\alpha^+_{0,0}\alpha^-_{m,n}c_{-m,-n}+
\tilde\alpha^+_{0,0}\tilde\alpha^-_{m,n}\tilde c_{-m,-n}\right),
\eqne
and it is the BRST-charge of an abelian theory. One may, as we will see below, use $Q_1$ 
to study the BRST-cohomology of the full theory. This requires us to determine the $Q_1$-cohomology, which 
we can do using the analysis of the abelian case given in the previous section. In order to apply this
analysis we need the existence of gauge fixing functions $\chi_{m,n}$ and $\tilde\chi_{m,n}$ such that
$\{\chi_{m,n},Q_1\}=c_{m,n}$ and $\{\tilde\chi_{m,n},Q_1\}=\tilde c_{m,n}$. Such functions exist if 
we assume that $\alpha^+_{0,0}$ and $\tilde\alpha^+_{0,0}$, which are conserved quantities, are
nonzero. Then $\chi_{m,n}=\frac{i}{m\alpha^+_{0,0}}\alpha^+_{m,n}$  for $m\neq 0$ and 
$\chi_{0,n}=\frac{1}{\alpha^+_{0,0}}q^+_n$ etc. for $\tilde\chi_{m,n}$. We will, 
in analyzing the cohomology, only consider functions that are finite degree polynomials in the basic 
fields, except $\alpha_{0,0}^+$ and $\tilde\alpha^+_{0,0}$,  where we permit inverse powers as well. 
Furthermore, we will assume no dependence on $q^-_0$ and $\tilde q^-_0$, which is sufficient for our 
case. 

We can now proceed and use the results of the previous section. This yields that the non-trivial 
polynomials in the cohomology of $Q_1$ have zero ghostnumber and have the dependence
\eqnb
h_{\{Q_1\}}&=&h_{\{Q_1\}}\left(q^I_n,\tilde q^I_n,q^-_{n\neq 0},\tilde q^-_{n\neq 0},\alpha^I_{m,n},\tilde\alpha^I_{m,n},\alpha^+_{0,n},
\tilde\alpha^+_{0,n}\right).
\eqne
where $I=1,\ldots,D-3$. Let us now study the cohomology of the string-like BRST-charge. 
One may expand a general BRST-invariant polynomial, $K$, in terms of its eigenvalues of $N_{lc}$ 
defined in 
eq.\ (\ref{N{lc}}) 
\eqnb
K&=&K_{N}+K_{N-1}+\ldots+K^I,
\label{K expansion}
\eqne
where
\eqnb
\{N_{lc},K_n\}&=&nK_n.
\eqne
By assumption, $N$ and $I$ are finite. The BRST-invariance implies
\eqnb
0=\{Q,K\}&=&\overbrace{\{Q_1,K_N\}}^{0}+\overbrace{\{Q_0,K_N\}+\{Q_1,K_{N-1}\}}^{0}+\ldots\no
&+&
\overbrace{\{Q_0,K_{N-i+1}\}+\{Q_1,K_{N-i}\}}^{0}+\ldots+\overbrace{\{Q_0,K_{I}\}}^{0},
\eqne
where $i=1,2,\ldots$. Thus, the highest order term, $K_N$, is BRST-invariant 
w.r.t.\ $Q_1$. Using the cohomology of $Q_1$, there exists two possibilities. 
Either $K$ has a ghost number different from zero which, by our analysis of the $Q_1$-cohomology, implies that the 
highest order term is BRST-trivial. This in turn implies, by the same reasoning, that all other terms 
with lower eigenvalue of $N_{lc}$ are trivial as well. Another possibility is that $K$ has zero ghost 
number. Although this case 
is not needed for our problem, we consider it out of general interest. 
For zero ghost number there can exist a non-trivial part in $K_N$
\eqnb
K_{N}&=&h^{(N)}_{N}+\{Q_1,C_{N-1}\},
\label{KN}
\eqne
where $h^{(N)}_{N}$ is a non-trivial function in the cohomology of $Q_1$ and $C_{N-1}$ has 
ghost number $-1$ and eigenvalue $(N-1)$ of $N_{lc}$. 
The phase-space function $h_{N}^{(N)}$ depends, by the analysis of the cohomology of $Q_1$, only on the 
fields 
$(q^I_n,\tilde q^I_n,q^-_{n\neq 0},\tilde q^-_{n\neq 0},\alpha^I_{m,n},$ $\tilde\alpha^I_{m,n},\alpha^+_{0,n},
\tilde\alpha^+_{0,n})$. Inserting eq.\ (\ref{KN}) into the 
equation for the next order yields
\eqnb
\{Q_0,h^{(N)}_{N}\}+\{Q_1,K_{N-1}-\{Q_0,C_{N-1}\}\}=0.
\label{K{N-1}}
\eqne
One can split $K_{N-1}$ into two parts $h_{N-1}^{(N)}+K'_{N-1}$ such that 
\eqnb
\{Q_1,h_{N-1}^{(N)}\}=-\{Q_0,h^{(N)}_{N}\}.
\eqne
This equation can always be solved since the right-hand side is $Q_1$-exact 
and has ghost number equal to one. Thus, from the $Q_1$-cohomology, there will always exist a function $h_{N-1}^{(N)}$. 
Eq.\ (\ref{K{N-1}}) now implies
\eqnb
\{Q_1,K'_{N-1}-\{Q_0,C_{N-1}\}\}=0,
\eqne
which is of a similar form as the equation previously solved. Consequently, the solution to $K_{N-1}$ 
is
\eqnb
K_{N-1}&=&h^{(N)}_{N-1}+h^{(N-1)}_{N-1}+\{Q_1,C_{(N-2)}\}+\{Q_0,C_{(N-1)}\},
\eqne
where $h^{(N-1)}_{N-1}$ is a function of $(q_n^I,\tilde q_n^I,q^-_{n\neq 0},\tilde q^-_{n\neq 0},\alpha^I_{m,n},\tilde\alpha^I_{m,n},\alpha^+_{0,n},
\tilde\alpha^+_{0,n})$. One can proceed in the same way to any order in $N_{lc}$. This yields the same kind 
of equations and the result in the end is
\eqnb
K&=&\sum_{i=I}^{N}\sum_{j=i}^{N} h^{(i)}_{j}+\{Q^0,C\},
\eqne
where we have defined 
\eqnb 
C&\equiv&\sum_{i=I}^{N} C_{i-1}.
\eqne
The functions $h_i^{(j)}$, where $i\leq N$ and $j\leq i$, are determined from the term with the highest eigenvalue of $N_{lc}$, thus, by $h_i^{(i)}$. This term only depends 
on the fields $(q_n^I,\tilde q_n^I,q^-_{n\neq 0},\tilde q^-_{n\neq 0},\alpha^I_{m,n},\tilde\alpha^I_{m,n},\alpha^+_{0,n},
\tilde\alpha^+_{0,n})$. Collecting the terms $h^{(j)}_{i}$, we can construct functions that are BRST-invariant 
and non-trivial w.r.t.\ the full string-like BRST-charge
\eqnb
h^{(j)}&\equiv&\sum_{i=-I}^j h^{(j)}_{i}.
\eqne
These functions are such that the term which has the highest value w.r.t.\ $N_{lc}$ is non-trivial in the 
cohomology of $Q_1$ and terms with lower eigenvalue, are correction terms such that the function is in 
the cohomology of $Q^0$.

Let us now conclude the analysis of the cohomology of the string-like BRST-charge. We have found that in the 
space of finite degree polynomials of the basic fields, excluding dependence on $q_0^-$ and $\tilde q_0^-$,
and assuming $\alpha_{0,0}^+$ and $\tilde\alpha_{0,0}^+$ to be non-zero as they enter in the 
expressions with 
inverse powers, the cohomology is non-trivial only for
zero ghost number. We may now use the result of section two to conclude that, provided our assumptions are valid, the
membrane BRST-charge is canonically equivalent to the string-like one.

Considering our assumptions, we have first of all the restriction to finite degree polynomials. 
This is always 
true within our perturbation theory.
Secondly, the assumption that $\alpha_{0,0}^+$ and $\tilde\alpha_{0,0}^+$ are non-zero is 
basically the same assumption 
one has for the known proof of the 
cohomology of string theory. As these fields are conserved quantities, this restricts possible initial 
conditions.  

The final assumption, namely the exclusion of 
$q_0^-$ and $\tilde q_0^-$ dependence, requires some more elaborate discussion. The zeroth and 
first order perturbation does 
not involve $q_0^-$ and $\tilde q_0^-$. This implies, by the proof of the cohomology in the abelian case in the 
previous section, that there 
exists an infinitesimal canonical transformation to first order. Using the gradation w.r.t.\ $N_{lc}$ 
defined in eq.\ (\ref{N{lc}}), one may construct the generator of the canonical transformation 
order by order in $N_{lc}$. It is straightforward to see that this generator will not depend on 
$q_0^-$ and $\tilde q_0^-$, which in turn implies that no higher order terms that are generated will 
depend on these fields, either. We can proceed in this way order by order proving the assertion.

%%%%%%%%%%%%%%%%%%%%%%%%%%%%%%%%%%%%%%%%%%%%%%%%%%%%%%%%%

\sect{Quantization}

We will in this section discuss the quantization of our model. This is done in the same manner  
as in the lightcone formulation in \cite{Bjornsson:2005rv}. We will, therefore, only repeat the 
essential features and discuss the differences of the two formulations. 

We have in the previous section proven that there exits, to any order in perturbation theory,
a canonical transformation connecting the stretched 
membrane model to the free string-like theory. We will now define the quantum theory for the 
stretched membrane from the free string-like theory by lifting 
the canonical transformations to unitary ones.  

We define the unitary transformations by an iterative procedure. At some arbitrary order $N$
in perturbation theory we define a unitary operator
\eqnb
U_N&\equiv&\exp\left(-i:_{N-1}G_N:_{N-1}\right).
\eqne
Here $G_N$ is the N'th order contribution to the generator of 
infinitesimal transformations, which we, from the previous section, know exists classically.
At the quantum level we specify the corresponding operator by the 
ordering $:_{N-1}$, which is the normal ordering 
w.r.t.\ the $(N-1)$'th order vacuum. This vacuum is defined 
by
\eqnb
\left| 0,k^+\neq 0\right>_{N-1}&=&U_{N-1}\cdot\ldots\cdot U_1 \left| 0,k^+\neq 0\right>_0,
\eqne
where the zeroth order vacuum, $\left| 0,k^+\neq 0\right>_0$, is defined in the usual way. Note that the
condition $k^+\neq 0$
is slightly different from the one in
the lightcone formulation in \cite{Bjornsson:2005rv}. The full unitary transformation to order $N$ in 
perturbation theory is then
\eqnb
U^{(N)}=U_N\cdot\ldots\cdot U_1.
\eqne

From the vacuum it is straightforward to construct the physical states for the stretched membrane theory 
to any finite order in perturbation theory. This is done in the same way as in the 
lightcone formulation. As an example, the new oscillators to order $N$ are defined as
\eqnb
\alpha_{m,n}^{(N),\mu}&\equiv&U_N\cdot\ldots\cdot U_1\alpha^\mu_{m,n}U^\dagger_1\cdot\ldots\cdot U^\dagger_N.
\eqne
Through our construction it follows immediately that to any order $N$ in perturbation theory
\eqnb
(Q)^2=\frac{1}{2}[Q, Q]=\frac{1}{2}[U^{(N)}Q^0{U^{(N)}}^\dagger, U^{(N)}Q^0{U^{(N)}}^\dagger]
=\frac{1}{2}[Q^0, Q^0]=0,
\eqne
where the last equality is true only for $D=27$. 

The partial gauge has singled out the $(D-1)$-direction and the corresponding field components
are given by
\eqnb
X^{D-1}&=&\frac{1}{\sqrt{g}}\xi^2\no
\mathcal P_{D-1}&=&-\sqrt{g}B, 
\eqne
where we have defined 
\eqnb
B&=&\mathcal P_\mu\partial_2 X^\mu. 
\eqne
One of the relevant physical operators found in \cite{Bjornsson:2005rv} involved the
integrated lightcone version of $B$. If we integrate $B$, denote it by $B_0$, then it is 
gauge invariant, but not BRST-invariant. In order to construct a BRST-invariant expression
one has to add ghosts to $B_0$. One will find the following 
BRST-invariant expression of $B_0$\footnote{
To get this expression we have redefined the ghost momenta by a factor $-i$,
such that one has the conventional anti-commutation relations}
\eqnb
B_0&=&\int d^2\xi\left\{\mathcal P_\mu\partial_2 X^\mu - i\partial_2 c b - i\partial_2 
\tilde c\tilde b\right\}.
\eqne
$B_0$ has the property that it is invariant under the constructed unitary transformations. 
This follows directly from the fact that $B_0$ is an eigenvalue operator which counts 
the mode number in the $\xi^2$-direction, and that the net the 
mode number of the unitary operators are zero. 

A final comment is that the BRST operator is only covariant w.\ r.\ t.\ the $D-1$-dimensional
subgroup of the full Lorentz group. Consequently, it is still an open question
whether the full Lorentz group is anomaly free.

%%%%%%%%%%%%%%%%%%%%%%%%%%%%%%%%%%%%%%%%%%%%%%%%%%%%%%%%%

\end{document}